\documentstyle[12pt]{article}
\input epsf
\newcommand{\be}{\begin{equation}}
\newcommand{\ee}{\end{equation}}
\newcommand{\ba}{\begin{eqnarray}}
\newcommand{\ea}{\end{eqnarray}}
\newcommand{\baa}{\begin{eqnarray*}}
\newcommand{\eaa}{\end{eqnarray*}}
\newcommand{\bb}{\begin{thebibliography}}
\newcommand{\eb}{\end{thebibliography}}

\newcommand{\bi}[1]{\bibitem{#1}}
\begin {document}
\begin{titlepage}
\begin{center}
{\Large \bf Proton-helium elastic scattering: a possible high-energy
polarimeter at RHIC-BNL}\\[1cm]
{C. Bourrely and J. Soffer}\\[0.3cm]
{\it Centre de Physique Th\'eorique - CNRS - Luminy,\\
Case 907 F-13288 Marseille Cedex 9 - France}
\end{center}
\begin{abstract}
We examine a suggestion to use $\mbox{p-} ^4\mbox{He}$ elastic scattering, as
an absolute polarimeter for high-energy polarized proton beams, by means 
of a Coulomb-Nuclear Interference effect for the single-spin asymmetry $A_N(t)$,
around the diffractive minimum of the differential cross section 
$|t|\sim 0.21 GeV^2$. Although this reaction has a fairly simple dynamical 
structure, our theoretical uncertainties and the present experimental
inaccuracy of the differential cross section in this $t$ region, allows one 
to generate dramatic effects for $A_N(t)$, which will be discussed.
\end{abstract}



\noindent Number of figures : 4

\smallskip

\noindent September 1998

\noindent Unit\'e Propre de Recherche 7061

\noindent CPT-98/P.3688

\noindent Web address: www.cpt.univ-mrs.fr
\end{titlepage}

\newpage
High-energy polarized proton beams are under construction at $RHIC-BNL$ and
thanks to the so-called {\it Siberian snake} technique, the beam polarization
is expected to reach the value $P=70\%$ and to be maintained at this impressive
high level. This is one of the key elements supporting the vast spin phenomena
programme for $pp$ collisions, which will be undertaken in the near future
\cite{RSC}. It has also motivated some detailed studies at $DESY$ in order to
decide whether or not
$HERA$ could operate as a $ep$ collider with both electron and proton beams
polarized \cite{HERA}. However one should have a reliable method for measuring
$P$ and the primary goal would be to achieve an accuracy of 5\% or better, {\it
i.e.} $\Delta P/P \leq 0.05$. This important calibration problem has given rise
recently to some activity, several methods have been proposed and their
limitations have been discussed \cite{BK}. In particular, one interesting
candidate is the so-called Coulomb-Nuclear Interference (CNI) polarimeter
relying on an idea, first suggested by Schwinger \cite{SCH}. 
If we consider $pp$
elastic scattering near the forward direction, say for $|t|\sim 10^{-3}GeV^2$,
the single-spin asymmetry $A_N(t)$ arises primarily from the interference
between the real electromagnetic helicity-flip amplitude and the imaginary
hadronic helicity-nonflip amplitude. It can be calculated exactly \cite{KL,BGL}
and one finds that it has a maximum value of about 4\% for $|t|\sim
3.10^{-3}GeV^2$,
which is almost energy independent. This effect has been investigated by the
E-704 experiment at $FNAL$ at $p_{lab}=200GeV/c$ \cite{AKC} and the results are
consistent with the theoretical prediction. However the situation is not as
simple as that, because the hadronic interaction need not conserve helicity in
the small $t$ region and the existence of a non-zero single-flip hadronic
amplitude introduces a substantial uncertainty on the predicted asymmetry \cite
{KZ,LT,BS}. Unfortunately, the lack of accuracy in the E-704 experiment leaves
too much freedom on the size of the single-flip 
hadronic amplitude and as a result,
given the present data, the CNI polarimeter is not a method which can achieve
the desired 5\% beam polarization error goal.

Another polarimetry method which involves $\mbox{p-} ^4\mbox{He}$ elastic
scattering has been
first briefly suggested in Ref. \cite{BK} and we think it deserves a careful
phenomenological analysis, which is presented in this paper. Since
$^4\mbox{He}$ is a spinless object, $\mbox{p-} ^4\mbox{He}$ elastic scattering
is a simple reaction which is described in terms of two helicity amplitudes,
the nonflip
$\phi_+(t)$ and the flip $\phi_-(t)$. The differential cross section reads
\be
{d\sigma(t) \over dt} =|\phi_+(t)|^2 + |\phi_-(t)|^2
\ee
and the single-spin asymmetry is
\be
A_N(t)={2Im[\phi_+(t)\phi_-(t)^*] \over |\phi_+(t)|^2 + |\phi_-(t)|^2 }~.
\ee
 $\phi_+$ and $\phi_-$ are written in terms of hadronic and electromagnetic
amplitudes in the form
$\phi_{\pm}(t)=\phi_{\pm}^h(t)+e^{i\delta}\phi_{\pm}^e(t)$, where $\delta$ is
the Coulomb phase shift. We have
\be
\phi_+^e(t)=-{4\alpha \sqrt{\pi}\over|t|}G_p(t)G_{He}(t)~,
\ee
where $\alpha$ is the fine-structure constant, $G_p(t)$ is the proton
electromagnetic form factor $G_p(t)=1/(1+|t|/0.71)^2$ and
$G_{He}(t)=[1-(2.56t)^6]e^{11.70t}$ is the $^4\mbox{He}$ electromagnetic form
factor \cite{JSC}.
Similarly we have
\be
\phi_-^e(t)=\sqrt{|t|}~{\mu_p -1 \over 2m_p}\phi_+^e(t)~,
\ee
where $\mu_p$ is the magnetic moment of the proton and $m_p$ its mass.

\begin{figure}
\begin{center}
\epsfxsize=13cm
\epsfysize=16cm
\centerline{\epsfbox{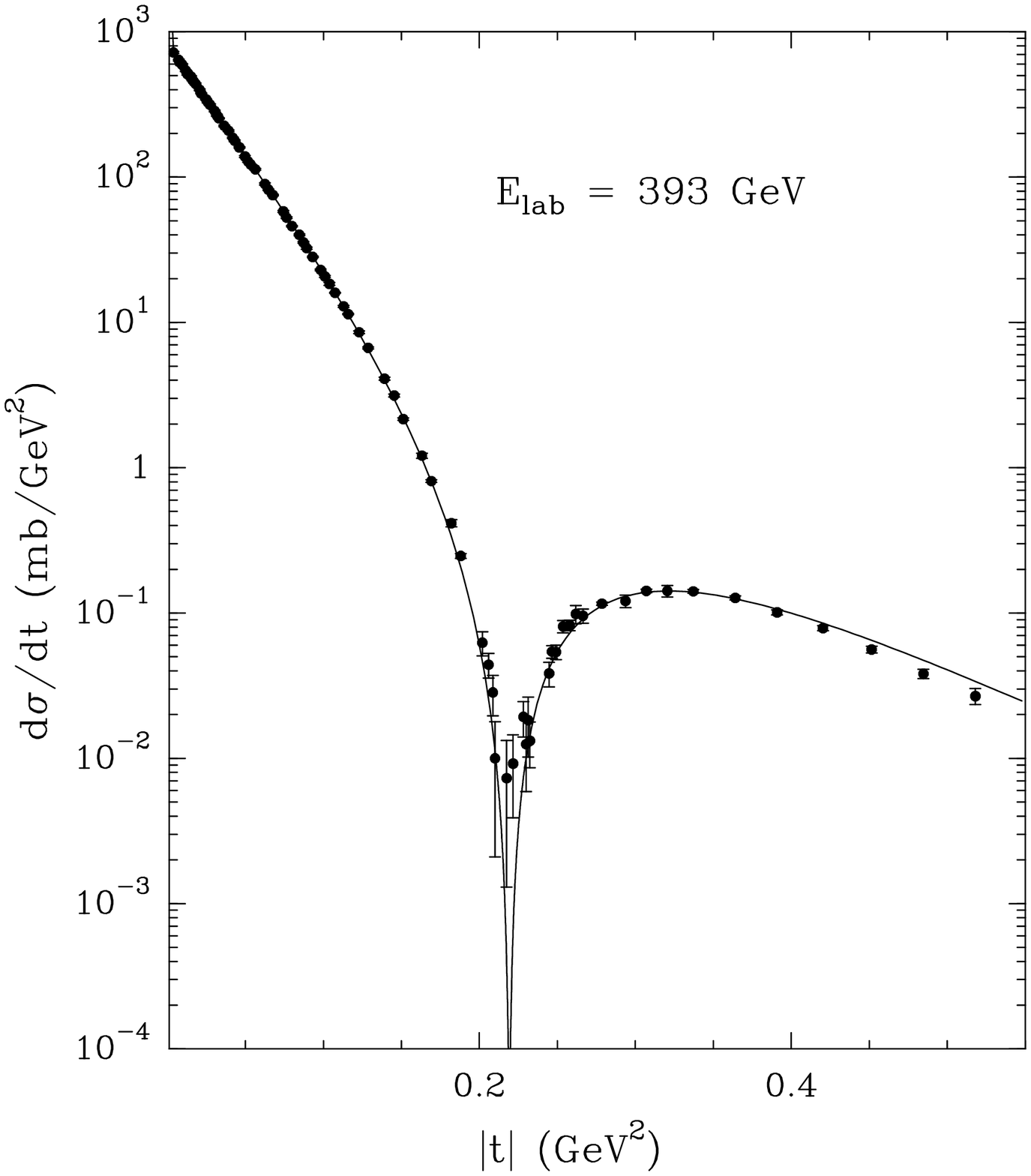}}
\caption{Differential cross section for $\mbox{p-} ^4\mbox{He}$ at
$E_{lab} = 393GeV$ as a function of $|t|$.}
\par{Data are from Ref.~\cite{data}.
The solid line is the result of our fit using eqs.(\ref{imphi}) and
(\ref{param}).}
\label{fi:dsig1}
\end{center}
\end{figure}

Our theoretical knowledge of $\phi_{\pm}^h(t)$ is less straightforward, but
before going into this discussion, let us briefly review the experimental
situation. At low energies, say, $1.1\leq p_{lab}\leq 2.5GeV/c$, the
differential cross section and the single-spin asymmetry have been accurately
measured at the $ZGS$-Argonne \cite{KLEM}, using polarized proton beams. The
cross section has a diffractive minimum around
$|t|=0.21GeV^2$ and $A_N(t)$, which is large (40-50\%), exhibits also an
interesting behavior in the dip region. At higher energies, say, $45\leq
p_{lab}\leq 400GeV/c$, only
$d\sigma/dt$ has been measured \cite{data}, and the diffractive minimum remains
essentially at the same $t$ value (see Fig.~\ref{fi:dsig1}).

\begin{figure}
\begin{center}
\epsfxsize=13cm
\epsfysize=16cm
\centerline{\epsfbox{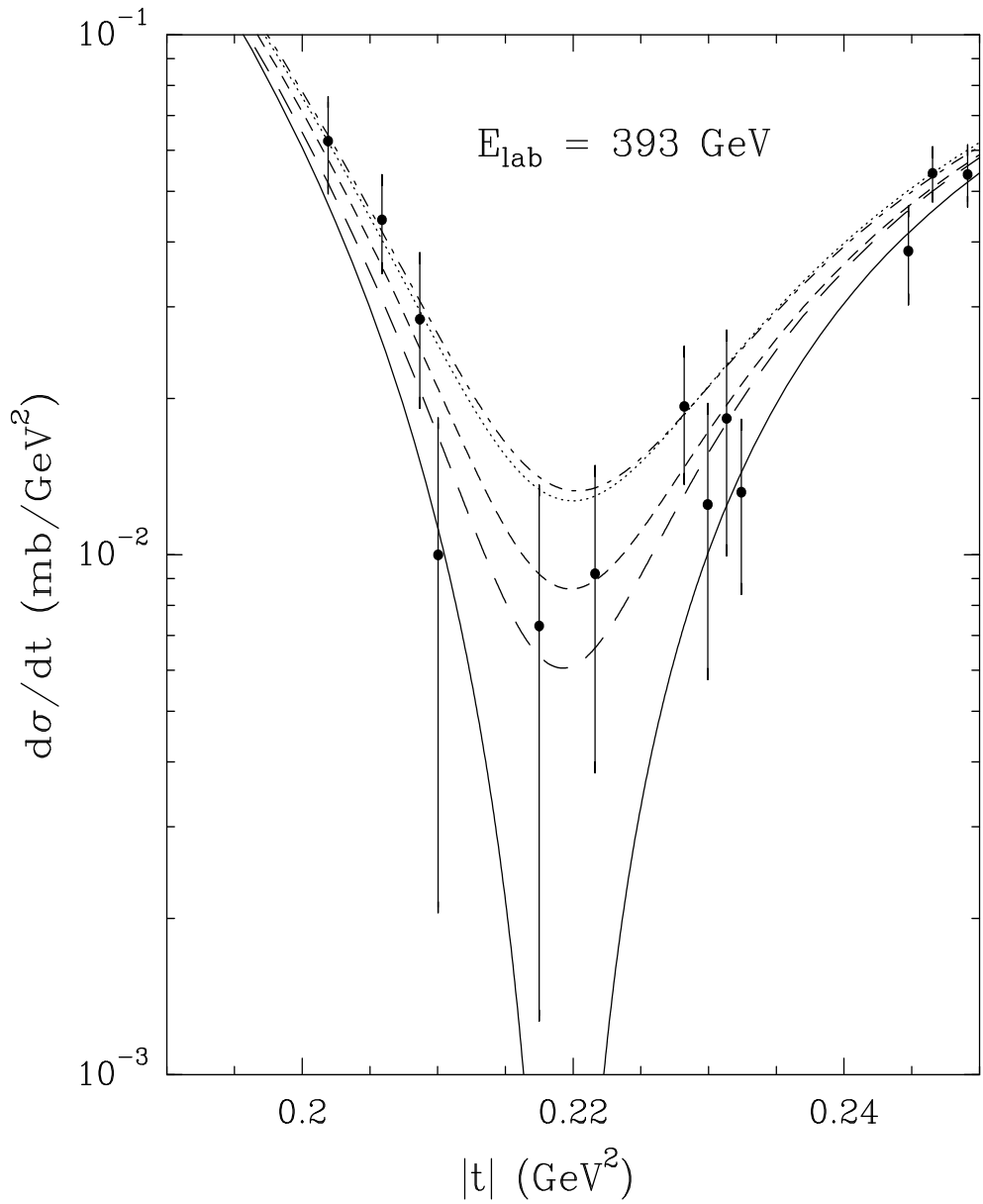}}
\caption{Enlarged dip region of Fig.~\ref{fi:dsig1}, showing different
possibilities described in the text.}
\label{fi:dsig3}
\end{center}
\end{figure}

\begin{figure}
\begin{center}
\epsfxsize=13cm
\epsfysize=16cm
\centerline{\epsfbox{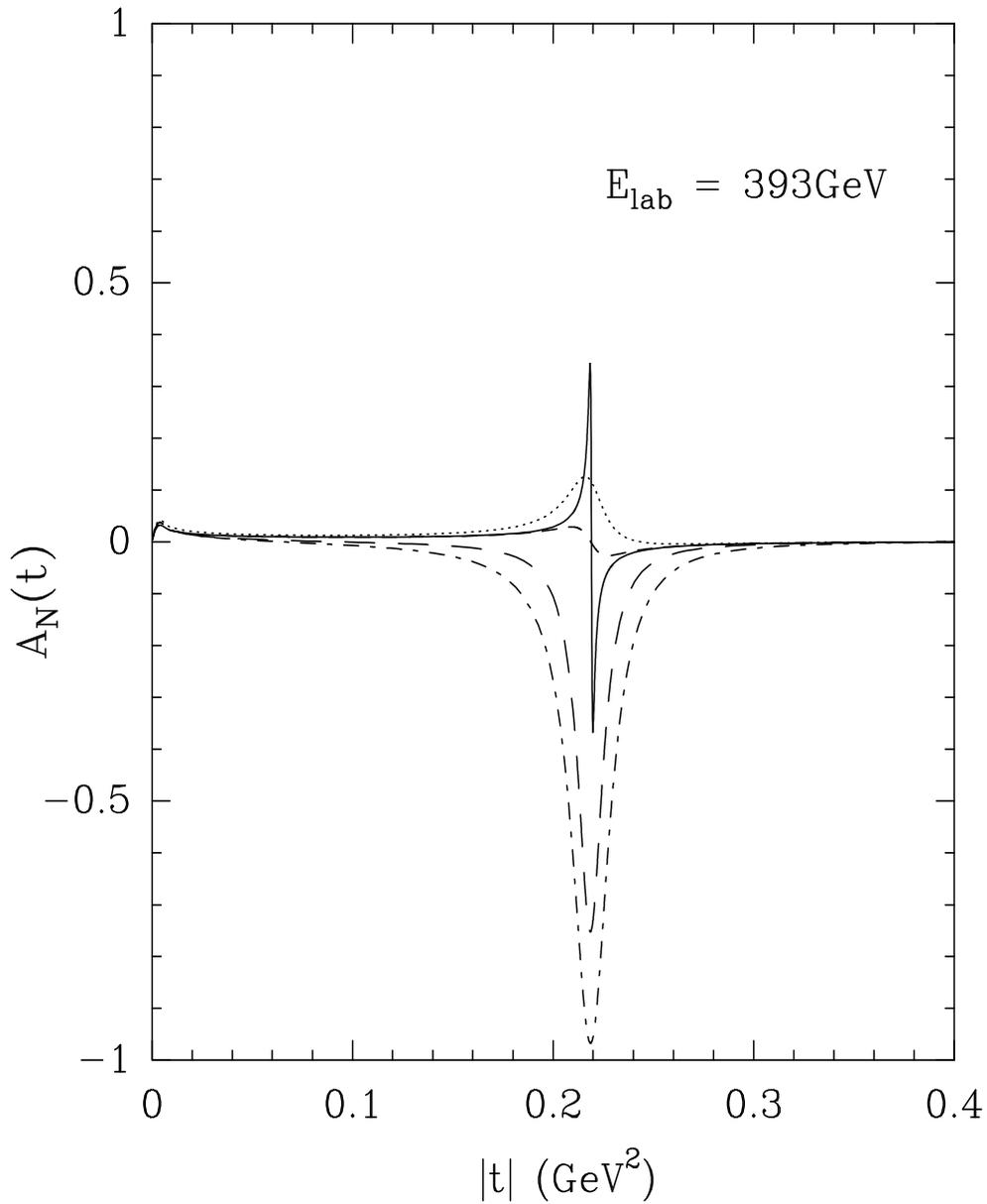}}
\caption{Single-spin asymmetry $A_N(t)$ for $\mbox{p-} ^4\mbox{He}$ at
$E_{lab} = 393GeV$ as
a function of $|t|$, showing different predictions explained in the text.}
\label{fi:an}
\end{center}
\end{figure}

\begin{figure}
\begin{center}
\epsfxsize=13cm
\epsfysize=16cm
\centerline{\epsfbox{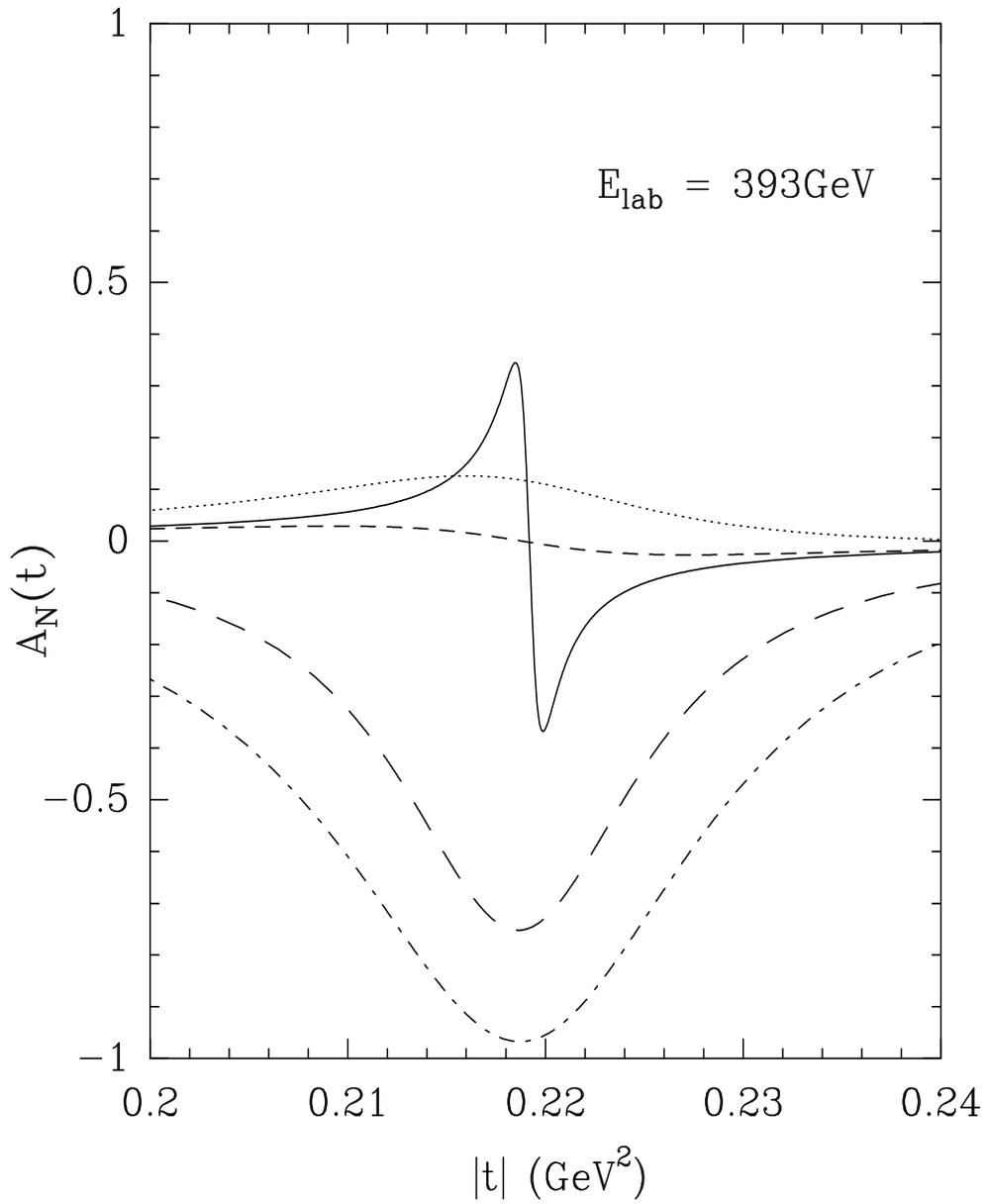}}
\caption{Enlarged dip region of Fig.~\ref{fi:an}.}
\label{fi:an1}
\end{center}
\end{figure}

Since we are concerned by the proton beams at $RHIC-BNL$, whose momentum lie
between $50GeV/c$ and $250GeV/c$, we will concentrate on the high-energy data.
Several analysis of these data have been made in the past based on, for
example, a Chou-Yang type model \cite{RJL} or a Glauber model \cite{data}, but
here we will present a rather simple phenomenological model. A Regge exchange
approach is greatly simplified by the fact that, since $^4\mbox{He}$ is an
isoscalar, the isovector "$\rho$ exchange", which has a large flip coupling, is
forbidden. Moreover only isoscalar trajectories  can be exchanged. They contribute mainly to
the non-flip amplitude $\phi_+(t)$ and the Pomeron prevails at very
high-energy. Consequently, one can assume the dominance of a purely 
diffractive Pomeron of the form
$Ae^{Bt}$, at fixed high-energy \footnote{The energy dependence and the phase
of the Pomeron, which have been obtained in a very successful analysis of $pp$
and $\bar pp$ elastic scattering \cite{BSW}, could be also used here, but it
goes beyond the scope of this paper.}. Therefore, as a first approximation, we
take the simple parametrization
\be
Im\phi_+^h(t)= Ae^{Bt} - Ce^{Dt}~,
\label{imphi}
\ee
where the second term stands for rescattering effects, so we expect $C \ll A$
and $D \ll B$. For the moment we neglect
$Re\phi_+^h(t)$ and $\phi_-^h(t)$, but we will come back to them later. The fit
of the cross section data at $E_{lab}=393GeV$, shown in
Figs.~\ref{fi:dsig1}-\ref{fi:dsig3}, is excellent and leads to the
following values of the parameters
\be
A=31.84\sqrt{mb}/GeV,~B=15.51GeV^{-2},~C=3.69\sqrt{mb}/GeV,~D=5.68GeV^{-2}~.
\label{param}
\ee
We have also well fitted the total cross section $\mbox{p-} ^4\mbox{He}$ value,
namely $\sigma_{tot}=(125.9 \pm 0.6)mb$. Note that in this case,
$Im\phi_+^h(t)$ changes sign at $|t|=0.219GeV^2$, so we find a very deep
diffractive minimum, namely $d\sigma/dt=5.10^{-5}mb/GeV^2$ at
this $|t|$ value, which is due to the contributions of $\phi^e_{\pm}(t)$.
Although
the data are not very accurate in this region, it would be surprizing to get
such a small cross section, but one cannot rule out such a possibility
\footnote  {Assuming a conservative $^4\mbox{He}$ jet density and a realistic
proton beam intensity, the luminosity is expected to be high enough, to allow
such a measurement with a reasonable accuracy (W.Guryn, private
communication).}. If we now calculate $A_N(t)$, it is driven by the product
$Im\phi_+^h(t)\cdot \phi^e_-(t)$ and
the result is depicted by the solid line in Fig.~\ref{fi:an}.
In the very small $|t|$ region we check that we have the usual CNI effect at
the level of 4\% or so, and in the vicinity of the dip, we find a strong
oscillation between +35\% and -35\%, which is better displayed in
Fig.~\ref{fi:an1}.

As already mentioned above, there is no fundamental theoretical reason to
believe that $\phi_-^h(t) =0$, even in a dynamical framework where the Pomeron
dominates. This important issue of the size of the Pomeron flip coupling has
been studied in details \cite{BNL,BKLST} and if we take, in analogy with
Eq.~(4),
\be
Im\phi_-^h(t)=r{\sqrt{|t|} \over m_p}Im \phi_+^h(t)~,
\ee
 one finds from different arguments a value of $r$ of 10\% or below. We have
also included such a contribution in our fit of $d\sigma/dt$ and the best fit
leads to $r=0.25$, with almost no changes in A, B, C and D. This new
contribution fills up the dip in the cross section, as show in
Fig.~\ref{fi:dsig3}
(dotted line) and we see in Figs.~\ref{fi:an}-\ref{fi:an1} that the strong
oscillation of $A_N(t)$ is now replaced by a smooth curve with a maximum value
of 13\% or so, in the dip region.

This situation is somehow oversimplified because, so far, we have neglected the
fact that from the data \cite{data}, one can extract $\rho$, the ratio of the
real to the imaginary part of the forward scattering amplitude and they find
$\rho=+0.102 \pm 0.035$. Therefore it is clear that we should take
$Re\phi_+^h(0)\not=0$. We don't know the $t$-dependence of $Re\phi_+^h(t)$
but for simplicity we will assume it has the slope B of the leading term
of $Im\phi_+^h(t)$. So if we now take
\be
Re\phi_+^h(t)=Ee^{Bt}~,
\ee
the best fit leads to $E=3.09\sqrt{mb}/GeV$. The net effect of this real part
is also to fill up the dip, as show in Fig.~\ref{fi:dsig3}.
We have first considered the case where $r=0$ (small dashed curve) and a second
case with $r\not=0$, which was fitted and led to $r=0.15$ (dotted-dashed
curve). This value is much smaller than that found above, in the absence of
$Re\phi_+^h(t)$. These two cases correspond to very different predictions for
$A_N(t)$ as shown in Figs.~\ref{fi:an}-\ref{fi:an1}. In the first case $A_N(t)$
is a smooth curve
(small dashed line) which changes sign at the dip position and in the second
case it
is very large and reaches almost -100\%. This effect is entirely due to the
product of $Re \phi_+^h(t)$ and $Im \phi_-^h(t)$, which dominate and become
almost equal in magnitude at the dip position. Since the sign of $Im
\phi_-^h(t)$ is unknown, by changing this sign one can have the mirror effect.
In this case, it is no longer a CNI effect.

Finally, one can envisage another realistic situation, where $Re \phi_+^h(t)$
has not the same t-dependence as $Im \phi_+^h(t)$. This is the case in various
models and in particular in Ref.~\cite{BSW}, where the real part decreases
faster than the imaginary part. So we have used again eq.(8) with a larger
value of the slope, that is $B=20GeV^{-2}$, and $r=0.15$. The results are shown
in Figs.~\ref{fi:dsig3}-\ref{fi:an}-\ref{fi:an1}, by the large dashed lines. As
expected, the filling of the dip is less pronounced than with the previous
value $B=15.51GeV^{-2}$. The shape of $A_N(t)$ is not affected, but its
magnitude is reduced accordingly.

To summarize, this phenomenological study of $\mbox{p-} ^4\mbox{He}$ elastic
scattering, shows that this simple reaction cannot be easily used as an
absolute polarimeter. An
accurate measurement of $d\sigma/dt$ in the dip region, might help us to pin
down the value of $Re\phi_+^h(t)$, but in order to clearly disentangle its
effect, on the filling of the dip, from that of $Im\phi_-^h(t)$, one certainly
needs
a direct measurement of $A_N(t)$, which hopefully, will have large 
values in the dip region.

We are grateful to Boris Kopeliovich and Tai Tsun Wu for several helpful
discussions.
\bb{99}
\bi{RSC} Proceedings of the Workshop {\it RHIC Spin Physics}, Riken BNL
Research
Center, April 27-29 (1998) (vol.7, Ed. T.D. Lee) report BNL-65615 and
references therein.
\bi{HERA} Proceedings of the Workshop {\it Deep Inelastic Scattering off
Polarized Targets:Theory meets Experiment}, DESY-Zeuthen, September 1-5 (1997)
(Eds. J. Bl\"umlein, A. De Roeck, T. Gehrmann and W.-D. Nowak) report DESY
97-200 and references therein.
\bi{BK} B.Z. Kopeliovich, High-Energy Polarimetry at RHIC, hep-ph/9801414.
\bi{SCH} J. Schwinger, Phys. Rev. {\bf 73}, 407 (1948).
\bi{KL} B.Z. Kopeliovich and I.I. Lapidus, Sov.J.Nucl. Phys. {\bf 19}, 114
(1974).
\bi{BGL} N.H. Buttimore, E. Gotsman and E. Leader, Phys. Rev. {\bf D18}, 694
(1978).
\bi{AKC} N. Akchurin {\it et al.}, Phys. Lett. {\bf 229B}, 299 (1989); Phys.
Rev. {\bf D48}, 3026 (1993).
\bi{KZ} B.Z. Kopeliovich and B.G. Zakharov, Phys. Lett. {\bf 226}, 156 (1989).
\bi{LT} L.T. Trueman, preprint BNL-63700, hep-ph/9610429.
\bi{BS} C. Bourrely and J. Soffer, Proceedings of the "12th Int. Symp. on
High-energy Spin Physics", Amsterdam Sept.10-14 (1996), World Scientific
(1997) p.825 (Eds. C.W. de Jager {\it et al.}).
\bi{JSC} J.S. Mc Carthy {\it et al.}, Phys. Rev. {\bf C15}, 1396 (1977).
\bi{KLEM} R. Klem {\it et al.}, Phys. Rev. Lett. {\bf 22},1272 (1977); Phys.
Lett. {\bf 70B}, 155 (1977).
\bi{data} A. Bujak {\it et al.}, Phys. Rev. {\bf D23}, 1895 (1981).
\bi{RJL} R.J. Lombard and A. Tellez-Arenas, Phys. Lett. {\bf 165B}, 205 (1985).
\bi{BSW} C. Bourrely, J. Soffer and T.T. Wu, Proceedings of the VIth Blois
Workshop, Blois, 20-24 June 1995, Editions Fronti\`eres 1996, p.15 and
references therein.
\bi{BNL} Proceedings of the Workshop {\it Hadron Spin-Flip at RHIC Energies},
Riken BNL Research Center, July 21-August 22 (1997) (vol.3, Ed. T.D. Lee)
report BNL-64724 and references therein.
\bi{BKLST} N.H. Buttimore, B.Z. Kopeliovich, E. Leader, J. Soffer and L.T.
Trueman, preprint CPT-98/P.3693 (in preparation).
\eb
\newpage

\newpage

\newpage

\newpage

\end{document}